\documentclass{article}

\usepackage{booktabs}
\usepackage{graphics}
\usepackage{graphicx}
\usepackage{listings}
\usepackage{nameref}
\usepackage{algorithm}
\usepackage{algpseudocode}
\algrenewcommand\algorithmicrequire{\textbf{Input:}}
\algrenewcommand\algorithmicensure{\textbf{Output:}}
\usepackage{bm}
\usepackage{amsmath}
\usepackage{amssymb}
\usepackage{caption} 
\usepackage{multirow}
\usepackage{color}
\usepackage[table]{xcolor}

\newcommand{\tool}{ViralQC}

\usepackage{arxiv}

\usepackage[utf8]{inputenc} 
\usepackage[T1]{fontenc}    
\usepackage{hyperref}       
\usepackage{url}            
\usepackage{booktabs}       
\usepackage{amsfonts}       
\usepackage{nicefrac}       
\usepackage{microtype}      
\usepackage{lipsum}
\usepackage{graphicx}
\graphicspath{ {./images/} }

\begin{document}
\title{ViralQC: A Tool for Assessing Completeness and Contamination of Predicted Viral Contigs}

\author{
 Cheng Peng \\
  Dept. of Electrical Engineering\\
  City University of Hong Kong\\
  Kowloon, Hong Kong SAR, China\\
  \And
 Jiayu Shang \\
  Dept. of Information Engineering\\
  Chinese University of Hong Kong\\
  Shatin, Hong Kong SAR, China\\
  \And
 Jiaojiao Guan \\
  Dept. of Electrical Engineering\\
  City University of Hong Kong\\
  Kowloon, Hong Kong SAR, China\\
    \And
 Yanni Sun \\
  Dept. of Electrical Engineering\\
  City University of Hong Kong\\
  Kowloon, Hong Kong SAR, China\\
}

\maketitle
\begin{abstract}
\textbf{Motivation:}  Viruses represent the most abundant biological entities on the planet and play vital roles in diverse ecosystems. Cataloging viruses across various environments is essential for understanding their properties and functions. Metagenomic sequencing has emerged as the most comprehensive method for virus discovery, enabling the sequencing of all genetic materials, including viruses, from host or environmental samples. However, distinguishing viral sequences from the vast background of cellular organism-derived reads in metagenomic data remains a significant challenge. While several learning-based tools, such as VirSorter2 and geNomad, have shown promise in identifying viral contigs, they often experience varying degrees of false positive rates due to noise in sequencing and assembly, shared genes between viruses and their hosts, and the formation of proviruses within host genomes. This highlights the urgent need for an accurate and efficient method to evaluate the quality of viral contigs. \\
\textbf{Results:} To address these challenges, we introduce \tool, a tool designed to assess the quality of reported viral contigs or bins. \tool\ identifies contamination regions within putative viral sequences using foundation models trained on viral and cellular genomes and estimates viral completeness through protein organization alignment. We evaluate \tool\ on multiple datasets and compare its performance against CheckV, the state-of-the-art in virus quality assessment. Notably, \tool\ correctly identifies 38\% more contamination than CheckV, while maintaining a median absolute error of only 3\%. In addition, \tool\ delivers more accurate results for medium- to high-quality (>50\% completeness) contigs, demonstrating its superior performance in completeness estimation.\\
\textbf{Availability:} The source code of \tool\ is available via: \href{https://github.com/ChengPENG-wolf/ViralQC}{https://github.com/ChengPENG-wolf/ViralQC}.\\
\textbf{Contact:} \href{yannisun@cityu.edu.hk}{yannisun@cityu.edu.hk}\\
\end{abstract}

\section{Introduction}
\label{sec:intro}

Viruses are the most abundant biological entities on Earth, playing crucial roles in diverse ecological systems, from human habitats to extreme environments \cite{mushegian2020there,cobian2016viruses,buscaglia2024adaptation}. To better understand their properties and functions, it is essential to catalog viruses across various environments. Metagenomic sequencing has emerged as the most comprehensive approach for viral discovery, enabling the sequencing of all genetic materials, including viruses, from host or environmental samples. For example, Istvan et al. \cite{istvan2024exploring} sequenced fecal samples from 1,034 individuals and identified over 18,000 DNA viruses. Rahlff et al. \cite{rahlff2023marine} investigated 55 metagenomes from marine, aerosol, and rain samples, uncovering viral exchange between rain and marine ecosystems.

While metagenomic sequencing is powerful in studying viruses, the generated sequencing data are highly heterogeneous, with virus-originated reads often obscured by the overwhelming presence of reads from cellular organisms \cite{dolja2018metagenomics,jochheim2024strain}. This makes the identification of virus-derived sequences within metagenomic data a significant challenge. To address this, numerous tools have been developed to identify viral contigs from assembled contigs. While alignment-based tools are efficient at identifying viruses that are closely related to known references, their ability to discover novel viruses is limited. Consequently, learning-based models have been increasingly incorporated into virus identification workflows \cite{ren2020identifying, guo2021virsorter2, camargo2024identification, peng2024viralm}. Although these learning-based tools differ in their training data, extracted features, and model architectures, they share the common goal of learning latent features beyond sequence similarity for more sensitive virus identification.

Despite the promising results of these virus identification tools, the quality of reported viral contigs remains a significant concern. Many of these contigs are fragmented and incomplete due to the inherent limitations in sequencing and assembly processes, resulting in the loss of critical genomic regions \cite{mallawaarachchi2023phables}. Furthermore, contamination from host genomes is a prevalent issue. Viral contigs frequently include sequences derived from host genomes, either due to shared genetic elements or the integration of proviruses into host genomes \cite{nayfach2021metagenomic}. These issues can hinder downstream analyses, such as functional annotation and evolutionary studies, distorting our understanding of viral communities.
Thus, understanding the quality of predicted contigs, including both contamination regions and completeness, is crucial for providing valuable insights to users. For instance, when conducting taxonomic classification for contigs and complete genomes, their closest matches based on sequence similarity may originate from different taxa, leading to wrong taxonomic classification for contigs from metagenomic data. By applying a completeness cutoff, it is possible to reduce errors in taxonomic assignment for contigs \cite{roux2019minimum,blanco2023extending}.

Assessing the quality of viral contigs poses several challenges due to the unique and diverse characteristics of viral genomes. Unlike cellular genomes, viruses exhibit high diversity in their genetic architectures, lacking universally conserved single-copy marker genes that are typically used for completeness estimation \cite{parks2015checkm}. Although sequence similarity is a common solution for finding the most related reference for completeness, it may fail on viruses that lack high similarity with existing reference genomes, especially at the nucleotide level.

Contamination detection in viral contigs introduces another layer of complexity. Existing approaches for contamination detection often rely on aligning sequences with reference genomes or databases. However, the limited coverage of current viral reference databases and the reliance on well-curated marker genes or sequence motifs can lead to errors in marking contamination regions in viral contigs from under-studied ecosystems.

\begin{figure}[h!]
    \centering
    \includegraphics[width=0.9\linewidth]{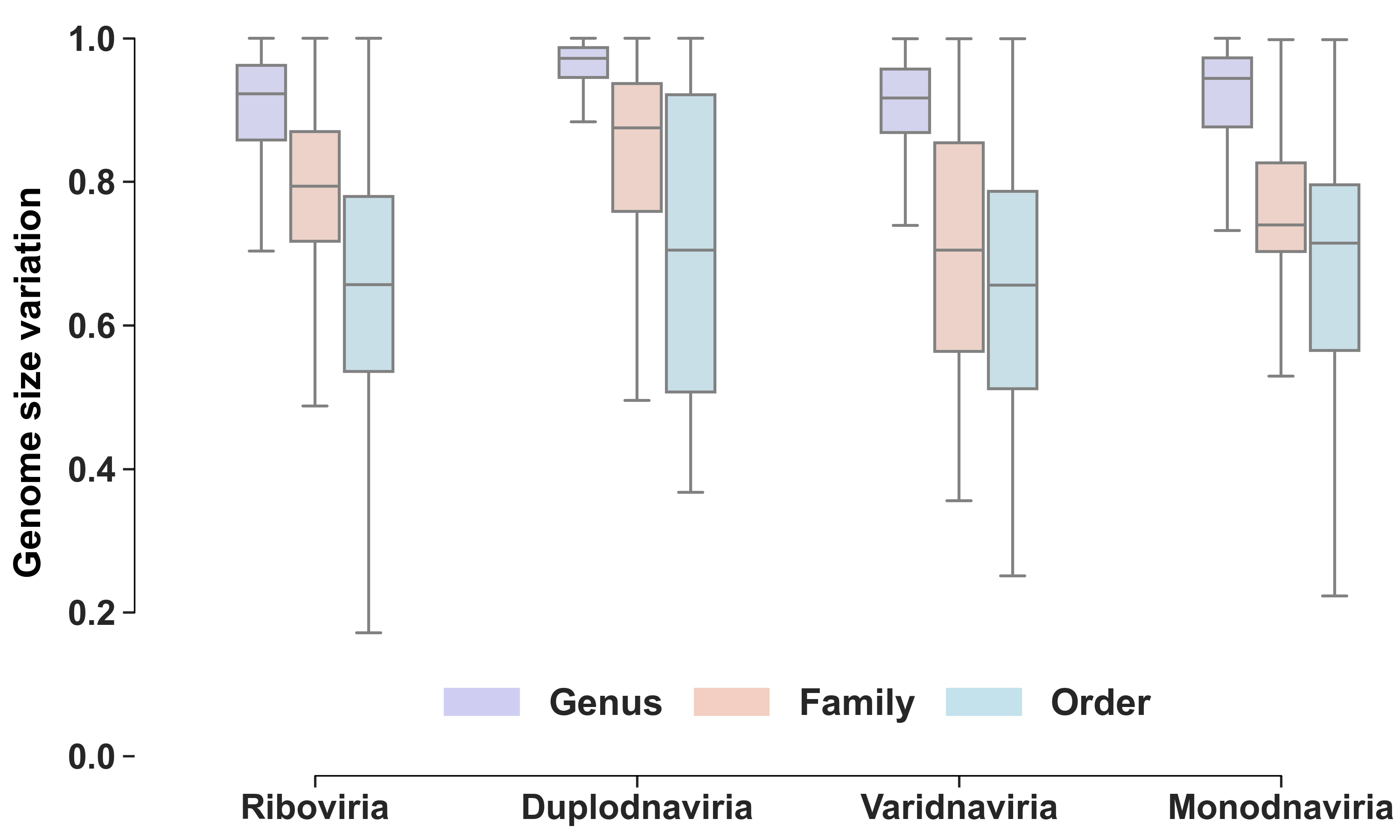}
    \caption{The distribution of viral genome size across taxonomic ranks. X-axis: viral realms. Y-axis: the ratio of genome length to the maximum genome length of the same taxonomic group. The increased ratio and reduced variance indicate greater genome size conservation from the order level down to the genus level across all realms.}
    \label{fig:violin_size}  
\end{figure}

\subsection{Overview}

\begin{figure}[h!]
    \centering
    \includegraphics[width=1\linewidth]{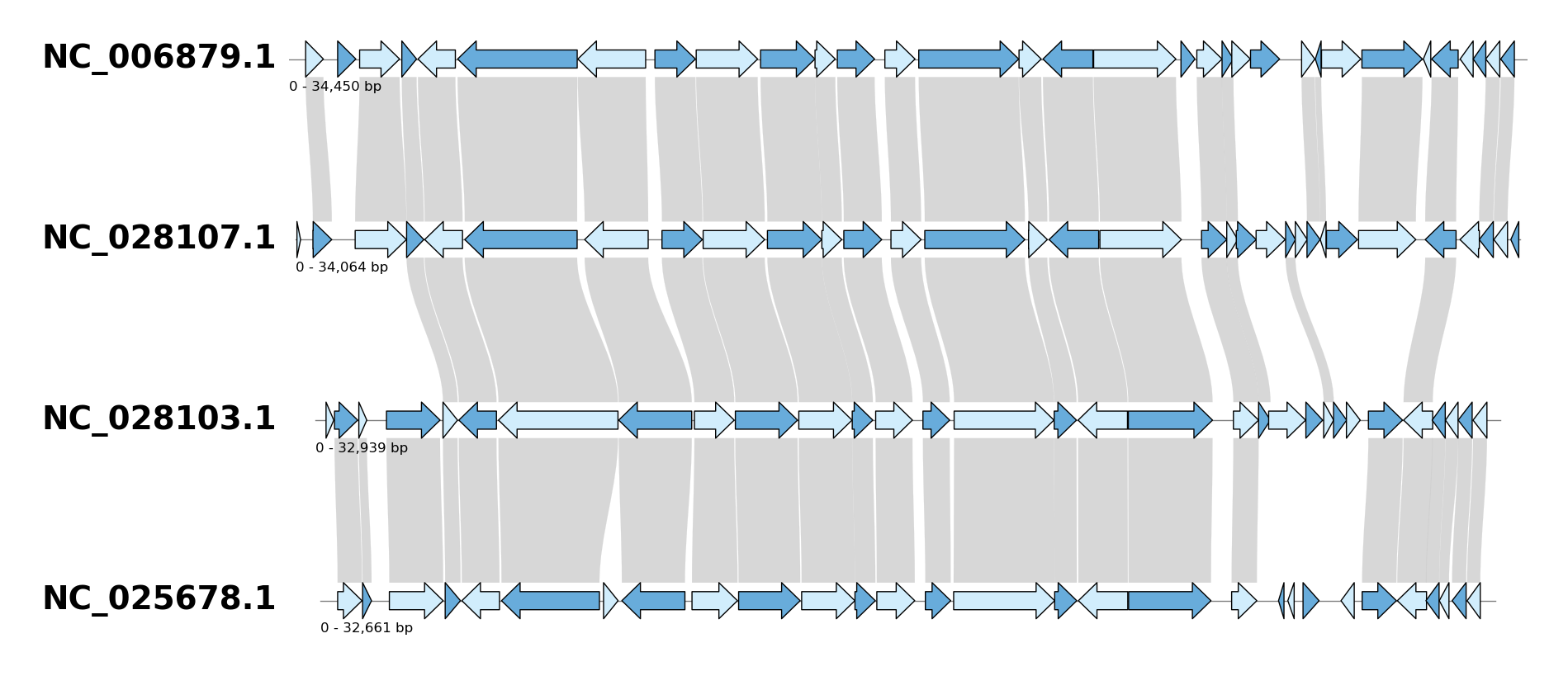}
    \caption{Example of conserved gene organization within genus \textit{Mastadenovirus}. The blue arrows represent proteins encoded in the genomes. The grey links indicate the similarity among the proteins.}
    \label{fig:synteny}  
\end{figure}

In this work, we present our new method \tool\ for assessing the completeness and contamination of predicted viral contigs. To address the challenges of high virus diversity, our new method identifies the best reference using two key observations. First, viral genomes tend to exhibit relatively consistent lengths within closely related taxonomic groups, particularly at the family or genus level. For example, viral genomes within the same genus show only a 6\% variation in length on average (Fig. \ref{fig:violin_size}). Second, related viruses often exhibit high conservation in the gene organization. As illustrated in Fig. \ref{fig:synteny}, genomes within the genus \textit{Mastadenovirus} exhibit similar gene organization, with homologous proteins arranged in a consistent order. These structural patterns reflect not only functional and evolutionary relevance, but also provide reliable indicators to assess genome completeness using genes/proteins.

Based on these observations, \tool\ is designed to automatically evaluate the quality of metagenome-assembled viral contigs by performing two critical tasks. First, it marks and reports the non-viral regions within viral contigs. To achieve this, \tool\ leverages the power of foundation models that are pre-trained on ~32 billion nucleotide bases and ~138 million proteins to enhance the detection of non-viral regions in a putative viral sequence. Second, \tool\ provides a quantitative estimate of viral contig completeness. Utilizing protein-based structural variations and sequence similarity, it identifies the most closely related virus, which provides a more accurate estimation of the complete genome length compared to current tools. 

We evaluated \tool's performance on multiple datasets, including a mock provirus dataset, a leave-one-taxon-out dataset, and metagenomic sequencing data. In addition, we benchmarked \tool\ against CheckV \cite{nayfach2021checkv}, which is the state-of-the-art tool for virus quality quantification. Our experiments demonstrate that \tool\ achieves high sensitivity and specificity in detecting contamination within viral sequences. Notably, \tool\ identifies 38\% more contamination than CheckV, while maintaining a median absolute error of only 3\%. In addition, \tool\ delivers more accurate results for medium- to high-quality (>50\% completeness) contigs, demonstrating its superior performance in completeness estimation.

\begin{figure*}[h!]
    \centering
    \includegraphics[width=0.95\linewidth]{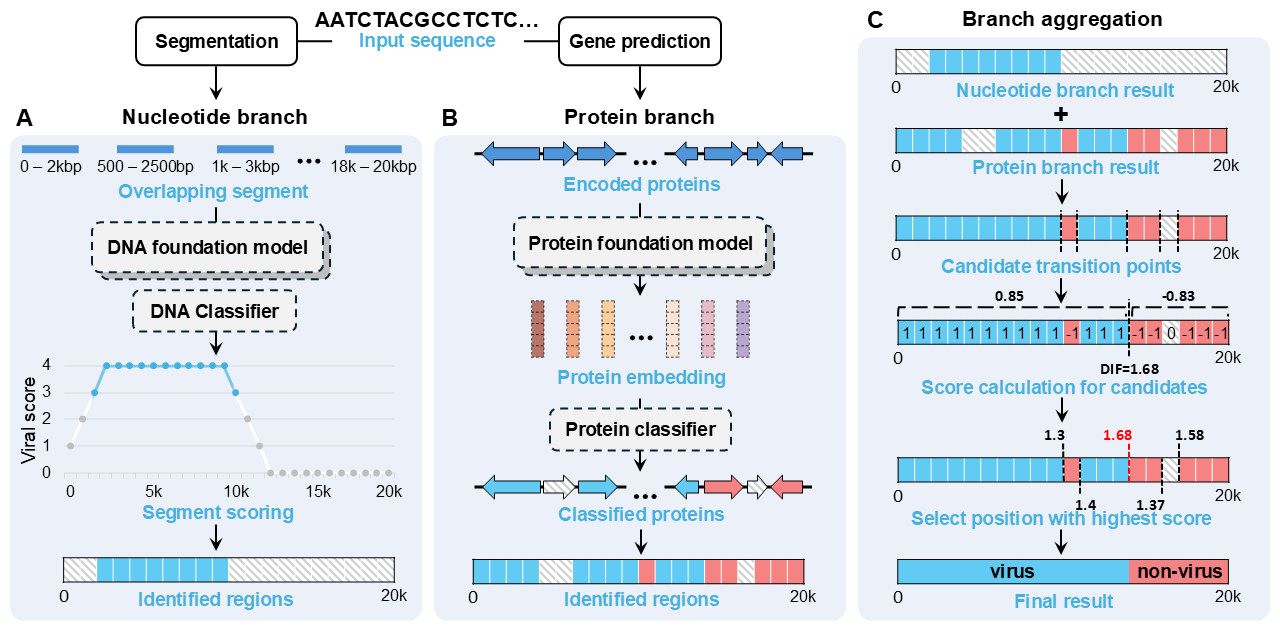}
    \caption{The framework of contamination detection module. (A) The nucleotide branch identifies viral regions based on sequence-level features using a fine-tuned DNA foundation model. (B) The protein branch classifies whether an encoded protein is virus-originated. (C) The results from both branches are aggregated and refined to produce the final predictions of virus and non-virus regions.
    }
    \label{fig:pipeline_contamination}
\end{figure*}

\section{Methods}
\label{sec:method}

\tool\ is organized into two primary modules. The contamination detection module (Fig. \ref{fig:pipeline_contamination}) identifies boundaries of non-viral regions in provided contigs, which often stem from assembled proviruses. The completeness estimation module (Fig. \ref{fig:pipeline_completeness}) evaluates the expected genome length and calculates the completeness of the viral contigs. Below we provide a detailed description of the methodology in each module.

\subsection{Contamination detection module}

To identify potential contamination in viral contigs, \tool\ employs a hybrid approach that leverages the complementary strengths of nucleotide and protein features. As discussed in Metabuli \cite{kim2024metabuli}, amino acid and nucleotide sequences have different sensitivity and specificity in sequence classification. While protein-based features are particularly useful for remote homology search, nucleotide-based features offer higher resolution for distinguishing between different taxonomic groups. By integrating both types of features, \tool\ enhances its ability on contamination detection.

The core of this module is foundation models, which are large language models pre-trained on extensive and diverse datasets, enabling them to be applied across a wide range of downstream tasks through fine-tuning \cite{zhou2024comprehensive}. One of the key advantages of using foundation models is their ability to learn rich, high-level representations of data, which allows them to generalize well to unseen sequences or rare events. This makes them particularly useful in complex biological tasks, where traditional methods might struggle to account for the vast diversity of genomic features \cite{li2024progress, rives2021biological, dalla2024nucleotide}. To mark the contamination region more accurately, we incorporate both DNA and protein foundation models in \tool. While the DNA foundation model specializes in capturing sequence-level features, the protein foundation model is able to provide supplementary information about whether a gene is likely from viruses.

Fig. \ref{fig:pipeline_contamination} sketches the main components of the contamination detection module. Given an input viral contig, \tool\ processes it through a nucleotide branch and a protein branch. The results provided by both branches are aggregated and reported, including the start and end positions of identified contamination regions. Below we describe the details of each branch. 

\subsubsection{Nucleotide branch}

The nucleotide branch is designed to classify viral regions based on the nucleotide makeup of sequences.
We leverage a fine-tuned DNA foundation model DNABERT-2 \cite{zhou2023dnabert} as the engine of the nucleotide branch to evaluate whether an input sequence is virus-originated. DNABERT-2 was pre-trained on a vast dataset including genomes from 135 species, enabling it to learn diverse and valuable representations of DNA sequences. These representations are particularly effective in distinguishing viral sequences from those of non-viral organisms. To adapt the DNA foundation model for viral region detection, we fine-tuned it to a binary classification task with two labels: viral sequences and non-viral sequences. By leveraging the pre-trained knowledge of the DNA foundation model and refining it for this specific task, the nucleotide branch achieves a powerful combination of generalization and specificity, making it highly effective in recognizing patterns and features specific to viral genomes.

The workflow of the nucleotide branch is illustrated in Fig. \ref{fig:pipeline_contamination} A. First, each contig is divided into overlapping 2kbp segments with a 500bp sliding step, ensuring that each 500bp region (except the ends of the contigs) is covered by four overlapping 2kbp segments. The fine-tuned DNA foundation model is adopted to decide whether each segment is either viral or non-viral. A viral score between 0 and 4 is assigned to each 500bp region, based on the number of covering segments predicted as viral. Regions with higher viral scores indicate stronger viral signals. Consecutive regions with viral scores greater than 2 are merged and labeled as viral regions. This overlap-based scoring approach effectively reduces false positives generated by learning models and ensures accurate detection of viral regions.

\begin{figure*}[h!]
    \centering
    \includegraphics[width=0.95\linewidth]{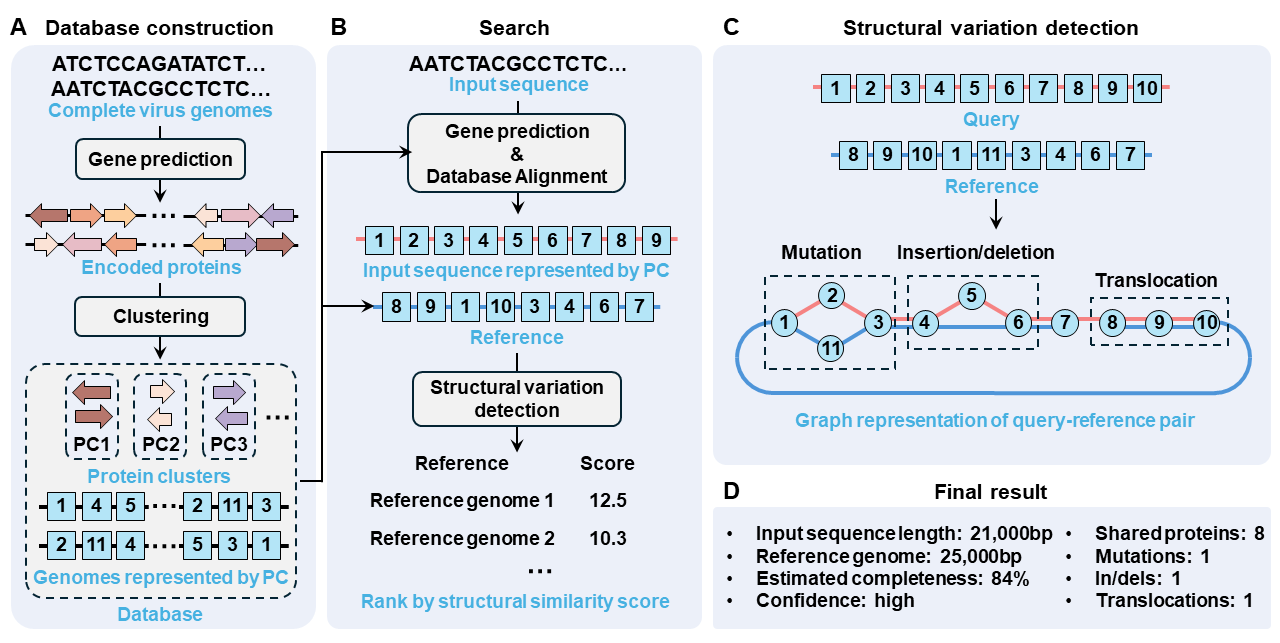}
    \caption{The framework of completeness estimation module. (A) Construction of the complete genome and protein cluster database. Proteins encoded by complete genomes are clustered into protein clusters. Each genome is represented by a series of protein clusters. (B) Homology search for the most related reference based on protein organization. Given an input sequence, the encoded proteins are predicted and aligned to the database. Then the input sequence is converted to a protein cluster sequence. Structural variations (SVs) in the sequence are detected and the structural similarity score is calculated. The references are ranked based on structural similarity score. (C) The detection of SVs using a graph-based method. Each query-reference pair is converted to a graph and the SVs are detected by analyzing the topology of the graph. (D) The completeness of the input sequence is estimated and reported based on the most related reference genome.}
    \label{fig:pipeline_completeness}  
\end{figure*}

\subsubsection{Protein branch}

The protein branch focuses on the prediction and classification of viral or non-viral proteins encoded by the input (Fig. \ref{fig:pipeline_contamination} B). It is designed to complement the nucleotide analysis by providing a higher-level understanding of sequence protein composition. This dual approach ensures that ambiguous or weak signals in the nucleotide branch are cross-validated with protein-based predictions.

To implement this, the protein branch utilizes the protein foundation model ESM2 \cite{lin2023evolutionary}, which is pre-trained on protein sequences from the UniRef database \cite{suzek2015uniref} using the Masked Language Modeling strategy \cite{devlin2018bert}. During pre-training, 15\% amino acids are masked, forcing the model to predict the masked residues and thus learn the underlying patterns and structures in the protein sequences. For each protein, ESM2 generates a per-residue embedding of $d_{l} \times d_{e}$, where $d_{l}$ is the number of residues in the protein sequence, and $d_{e}$ is the embedding size. The embedding is a meaningful representation of the protein sequence that encapsulates sequence composition and biological structure in a high-dimensional space, making them suitable for classifying proteins.

The protein branch begins by predicting proteins from input contigs using Prodigal-gv \cite{camargo2024identification}, a modified version of Prodigal \cite{hyatt2010prodigal} that optimizes gene calling for giant viruses and viruses that use alternative genetic codes. Then the predicted amino acid sequences are encoded into per-residue embeddings using ESM2. Protein embeddings are obtained by averaging over all residues to reduce computational complexity. Based on these protein embeddings, the binary classifier computes a score between 0 to 1 for each protein, indicating the likelihood of the protein being a viral protein. Stringent thresholds are employed to eliminate false positives and false negatives. Specifically, proteins with scores above a threshold (0.5) are classified as \textit{viral proteins}, while those below a threshold (0.1) are classified as \textit{non-viral}. Proteins falling between these thresholds refer to ambiguous cases and are classified as \textit{unknown}.

\subsubsection{Branch aggregation}

After obtaining results from both the protein and nucleotide branches, \tool\ aggregates these findings to predict the contamination regions. First, the protein classification results are refined using the viral regions detected by the nucleotide branch. Since our tests show that the nucleotide branch has high specificity for virus detection, its predicted viral regions are used to re-label proteins initially classified as \textit{unknown} or \textit{non-viral} by the protein branch. Specifically, if more than 50\% of an unknown/non-viral protein’s length overlaps with a viral region identified by the nucleotide branch, it is reclassified as a viral protein. After this refinement, contiguous regions with the same classification are merged to form larger, unified regions.

Second, as shown in Fig. \ref{fig:pipeline_contamination} C, small fragmented regions may occur when viral and non-viral regions are not clearly delineated, particularly in cases where genes with conflicting labels from the DNA and protein branches are scattered across the contig. To address this and remove potentially false predictions, we refine the results by searching for \textit{transition points}, which represent positions where the gene content changes significantly.

Specifically, candidate \textit{transition points} are determined as loci between viral and non-viral regions identified in the first step to narrow the searching space. Each protein is assigned a score based on its label (+1 for viral, 0 for unknown, and -1 for non-viral). \tool\ scans through the candidate positions within the contig. For each candidate position, the average scores of the left side (from the start of the contig to the candidate position) and the right side (from the candidate position to a position that includes at least 10\% genes) of the contig are calculated respectively. A candidate is identified as a \textit{transition point} if it has the maximum absolute score difference between the two sides among all candidates and the score difference exceeds a pre-defined threshold. The threshold is set to 1 to ensure significant gene content differences between the two sides. Once a \textit{transition point} is identified, the region to the left of the transition point is classified as viral/non-viral based on its gene content and excluded from further analysis. The above steps are repeated on the remaining portion of the contig until no additional \textit{transition points} are identified. The \textit{transition point} step identifies the true virus-non-virus boundaries and prevents the contigs from over-fragmentation, improving \tool's robustness in complex contigs. An example of this process is illustrated in Fig. \ref{fig:pipeline_contamination} C.

\subsection{Completeness estimation module}

Building on the observed consistency of genome length and conservation of gene structure within related taxonomic groups, \tool\ estimates the completeness of an input viral contig in two steps. First, it identifies the most closely related reference genome from a complete viral genome database based on protein structural similarities (Fig. \ref{fig:pipeline_completeness} B). Then, the completeness is calculated as the ratio of the input contig length to the length of the identified reference genome, which is a value ranging from 0 to 1 (Fig. \ref{fig:pipeline_completeness} D). 

In addition to single contigs, \tool\ also supports completeness estimation for entire bins using the ``--bin'' option. Users are not required to manually concatenate contigs within a bin before running \tool. This function enables the evaluation of bin completeness as well as the analysis of segmented viruses. The key difference between single-contig mode and bin mode is that, in the single-contig mode, \tool\ treats each segment of a segmented virus as a separate complete genome. In contrast, in bin mode, all segments are considered part of a single genome. This flexibility allows users to evaluate both individual segments and entire segmented viruses efficiently.

\subsubsection{Homology search incorporating protein organization}
To accurately estimate the completeness of a viral contig, it is essential to identify the most closely related reference genome. Traditional approaches often rely solely on sequence similarity, which can be insufficient for highly divergent or novel viruses. To overcome these limitations, \tool\ incorporates not only sequence similarity but also the structural organization of proteins.

\paragraph{Protein cluster}
To incorporate both sequence similarity and protein organization, we construct protein clusters using complete viral genomes from NCBI RefSeq. 
First, protein sequences predicted by Prodigal-gv are clustered into protein clusters (PCs) at 50\% identity and 80\% coverage using MMseqs2 \cite{steinegger2017mmseqs2}. Proteins within the same cluster are considered homologous and perform similar biological functions. Each cluster is assigned a unique PC ID. With proteins labeled by their corresponding PC ID, each genome is represented as a series of protein clusters, indicating both the protein composition and the order of proteins within the genome. Examples of such digital representations of protein organization are illustrated in Fig. \ref{fig:pipeline_completeness} A.

\paragraph{Structural variation} To assess structural similarities at the protein level, \tool\ identifies four types of structural variations (SVs) between two sequences: mutations, insertions/deletions, translocations, and duplications. Mutations arise from the accumulation of nucleotide variations within a gene that changes the homologous family of the protein. Insertions involve the acquisition of external genes from other viruses or hosts, while deletions are the loss of genes. Translocations refer to the rearrangement of DNA fragments within the genome that changes genes' positions. Duplications lead to differences in gene copy numbers. These events induce differences in gene organization, which are considered to have an opposite effect on structural similarity. Fewer SVs suggest that two sequences share more similar gene organization and are therefore more evolutionarily related.

\paragraph{SV detection using graph algorithms}
Given an input viral contig, \tool\ first conducts gene prediction and translation using Prodigal-gv. The predicted proteins are aligned to the proteins in the complete viral genome database using DIAMOND v2.1.9 \cite{buchfink2021sensitive}. The ``ultra-sensitive'' option is specified to enhance the detection of distantly homologous proteins while maintaining computational efficiency. Based on the alignment results, the query proteins are labeled by the PC IDs of their best hits, and the input contig is represented by a series of proteins in the same way as the complete genomes in the database.

\tool\ utilizes a graph-based method to detect SVs between the query and reference genomes. In this approach, both sequences are converted into a graph where nodes represent proteins in the sequences and an edge exists between two proteins if they are adjacent in any of the sequences. Proteins shared between the query and reference (with the same PC ID) are represented by the same node. Since query contigs are often incomplete and align to only part of a reference genome, \tool\ focuses on the aligned region rather than the entire genome. Proteins in the reference genome outside the aligned region are excluded, while query proteins outside the region are labeled as outliers and penalized during scoring.

To detect SVs, \tool\ analyzes the topology of the constructed graph. Duplications can be easily identified by examining the number of proteins with identical PC IDs within the same sequence. Sequence-specific nodes that present in only one of the two sequences may arise from mutations, insertions, or deletions. Specifically, \tool\ scans the whole graph and records all unique nodes and their neighbors. Then, mutations are defined as pairs of unique nodes that share the same neighbors. Insertions refer to nodes that appear only in the query contig, while deletions are nodes unique to the reference genome. Translocation events are more difficult to estimate directly. Instead, \tool\ divides the graph into subgraphs, referred to as synteny blocks. Proteins within a synteny block are conserved in order, but the order of synteny blocks themselves may vary. The number of synteny blocks acts as an indicator of translocations. Fig. \ref{fig:pipeline_completeness} C illustrates an example of these SVs displayed in the genome graph. 

To determine the most closely related reference genome, \tool\ calculates a structural similarity score that incorporates both shared proteins and observed structural variations. The underlying assumption is that closely related reference genomes will share a higher number of proteins while exhibiting fewer structural variations. As a result, the structural similarity score is computed as the weighted sum of two components: the number of shared proteins and the number of observed SVs. Shared proteins contribute positively to the score, while SVs are penalized to account for deviations from the expected structural arrangement. The reference with the highest structural similarity score is selected as the most related genome and used to compute the completeness. In cases where multiple references achieve the same similarity score, or when the query contig is highly divergent (e.g., sharing few genes and exhibiting many SVs), the structural similarity approach may not be effective. In such cases, \tool\ selects the reference genome with the highest overall protein alignment score, calculated as the sum of the bit-scores of aligned proteins.

\subsubsection{Confidence level}

While homolog search based on gene organization effectively identifies evolutionary-related genomes, false positives can occur when the query viral contig is highly divergent to the database (e.g., share only a few proteins with the best hit). To address this, \tool\ reports a confidence level for each result, based on the number of shared proteins and genome size.
Inspired by vConTACT \cite{bolduc2017vcontact}, we assume that all $n$ protein clusters have an equal probability of being chosen. The probability that two genomes with $a$ and $b$ protein clusters, respectively, share at least $c$ clusters by chance is computed in Eqn. \ref{eq:pval}. Then Eqn. \ref{eq:sig} calculates the statistical significance score of query-reference pairs with $c$ shared proteins out of the database size $T$. Based on the significance score, \tool\ classify the results into three confidence levels: high confidence (score $\ge$ 5), medium confidence (5 < score $\le$ 1), and low confidence (score $\le$ 1).

\begin{equation}
\label{eq:pval}
P(x\ge c)=\sum_{i=c}^{min(a,b)}\frac{C_{a}^{i} C_{n-a}^{b-i}}{C_{n}^{b}}
\end{equation}

\begin{equation}
\label{eq:sig}
Confidence(A,B)=-log(P\times T)
\end{equation}

\section{Result}
\label{sec:result}

\tool\ takes viral contigs as input. It first identifies and marks potential contamination regions within the contigs. Then it removes these regions and provides an evaluation of genome completeness. Since the methodologies underlying contamination detection and completeness estimation differ - contamination detection is based on deep learning, while completeness estimation is based on structure-aware homology search - we present their results separately. In the following sections, we will first evaluate the performance in contamination detection, followed by completeness estimation. Lastly, we will present a case study demonstrating the application of \tool\ on metagenome-assembled contigs derived from a soil sequencing dataset.

\subsection{Contamination detection}

\subsubsection{Training and test data}
To evaluate the performance of contamination detection, we constructed a mock provirus dataset by combining viral and cellular genomes. 
First, we downloaded viral genomes released before September 2023 from the NCBI RefSeq database (https://www.ncbi.nlm.nih.gov/), along with a set of high-quality metagenomic viral sequences from previous studies \cite{guo2021virsorter2}. These datasets include genomes from both known and novel viruses, resulting in a total of 49,929 viral genomes, which were used as positive training samples. To construct negative samples, we included bacterial and archaeal assemblies that serve as hosts of prokaryotic viruses from the NCBI RefSeq database. To alleviate the huge class imbalance between positive and negative samples, we down-sampled the negative dataset by randomly selecting two genomes per genus, resulting in 1,651 assemblies. Then, the combined dataset was randomly divided into training and test sets with a ratio of nine to one. To address the need for contamination detection in contigs, all genomes were fragmented into contigs of varying lengths. These fragmented contigs were used as the training set for the DNA branch of the contamination detection module.

For the protein branch, proteins in the training set were predicted using Prodigal-gv. All predicted proteins were clustered using MMseq2 with an identity threshold of >50\% and coverage >80\%. Protein clusters with a purely viral origin and those with a purely cellular origin were used as positive and negative training data, respectively. Protein clusters shared by both origins were excluded from model training to avoid introducing ambiguity into the model.

To evaluate contamination detection performance, a test set of mock provirus sequences was constructed. Viral genomes in the test set were randomly fragmented into segments of varying lengths, which were then inserted at random positions within bacterial and archaeal genome fragments to simulate provirus sequences These simulated sequences included varying lengths (5k, 10b, 20b, and 50kbp) and different contamination levels (10\%, 20\%, and 50\%).

\paragraph{Evaluation metric} We adopt ``absolute error'' ${E}_{cont}$ to evaluate the performance of \tool\ in contamination detection. Each base in an input contig belongs to either a viral or non-viral (i.e., contamination) region. Thus, ${E}_{cont}$ is the percentage of misclassified bases (i.e., $n$) in a contig of length $l$, as shown below.

\begin{equation}
    \label{m1}
    {E}_{cont} = \frac{n}{l} 
\end{equation}

We validate \tool\ on several datasets and compare its performance with CheckV (v1.0.3) \cite{nayfach2021checkv}. Because CheckV uses a curated marker gene database to identify virus-host boundaries, we directly run CheckV with its v1.5 database for contamination detection.

\begin{figure}[h!]
    \centering
    \includegraphics[width=0.9\linewidth]{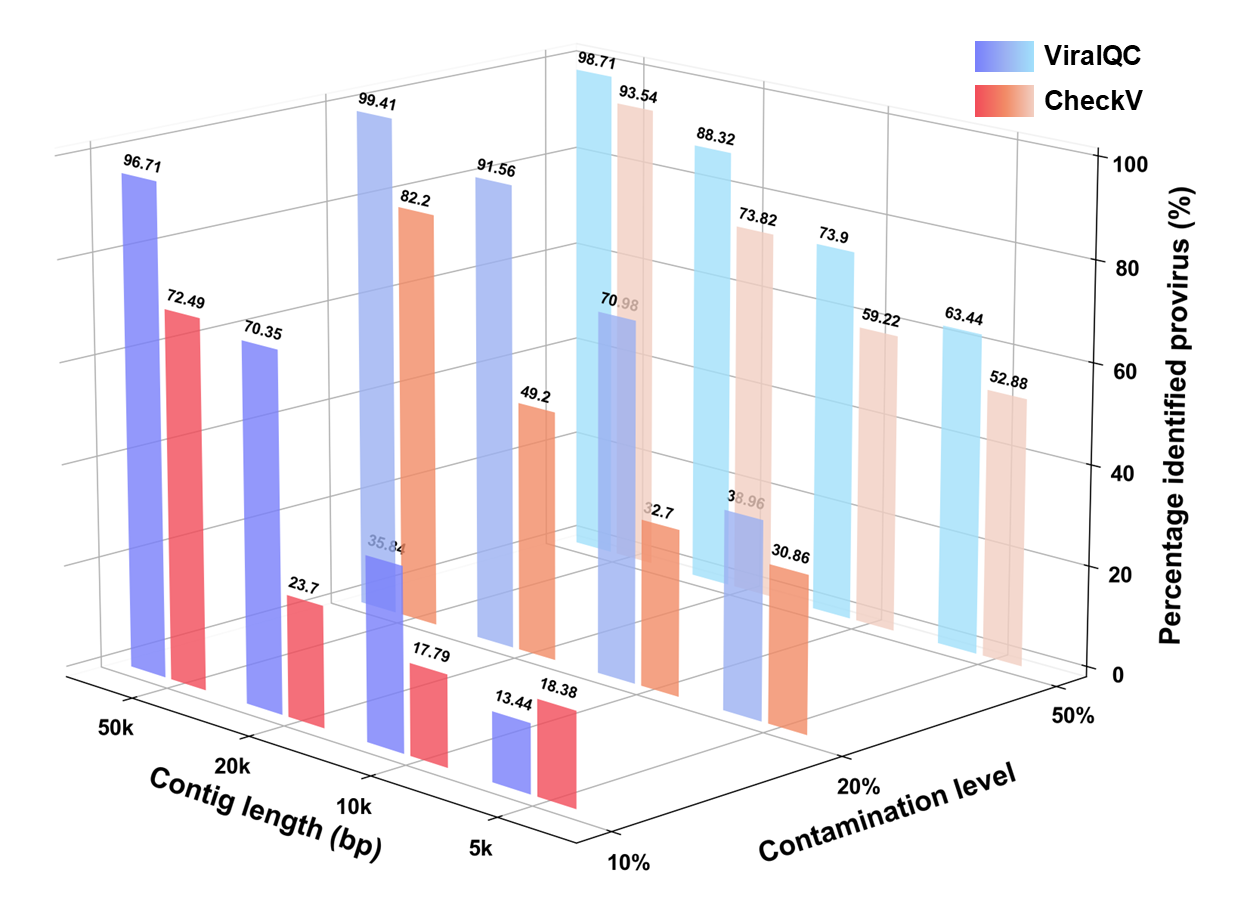}
    \caption{Percentage of identified provirus sequences on contigs with varying lengths (5k-50kbp) and contamination levels (10\%-50\%). As all the inputs contain contamination, the Z-axis represents the rate of contamination discovery.}
    \label{fig:contamination_predict_num}  
\end{figure}

\begin{figure}[b]
    \centering
    \includegraphics[width=\linewidth]{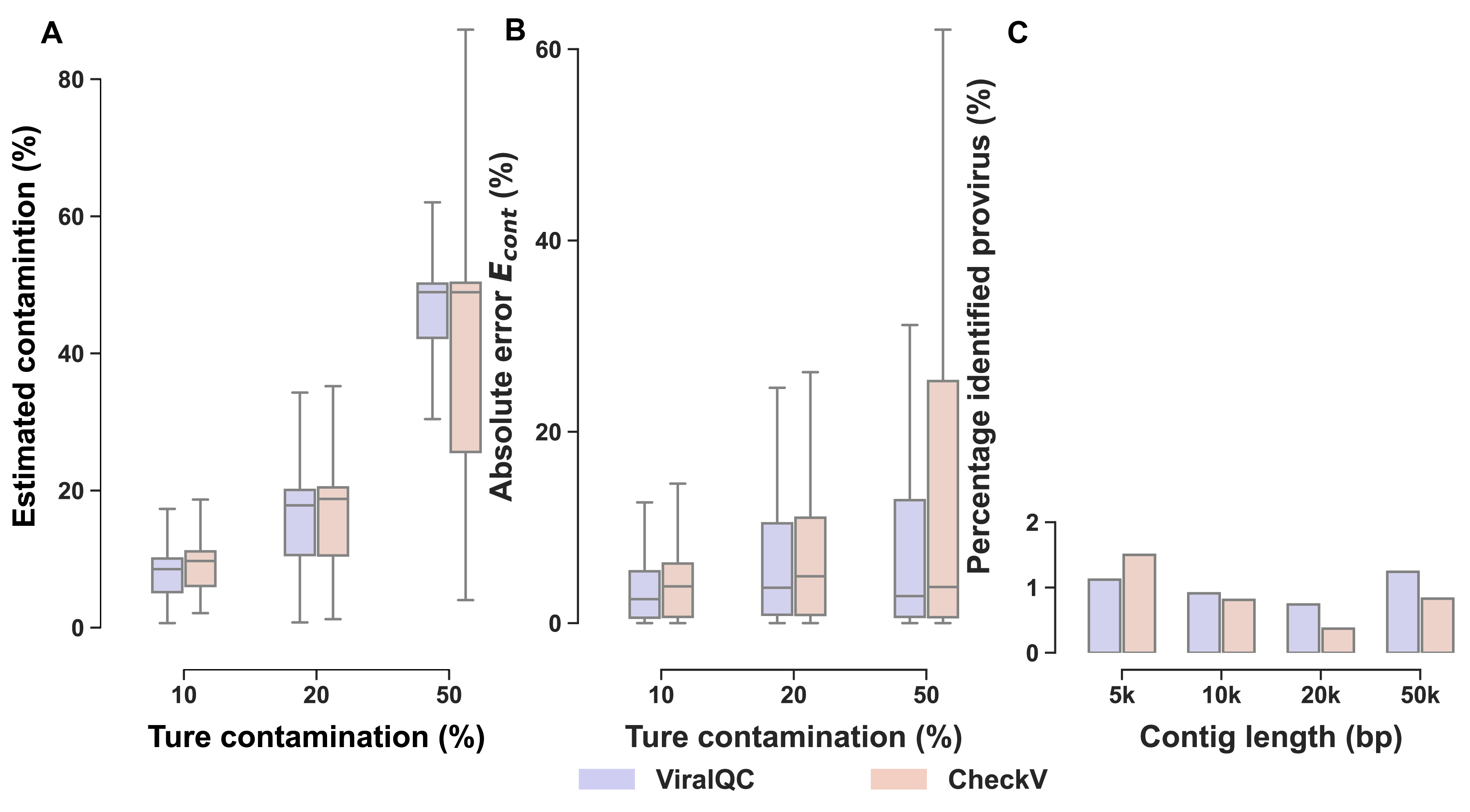}
    \caption{(A) The distribution of estimated contamination against true contamination for identified provirus sequences on contigs with varying contamination levels (10\%-50\%). (B) The distribution of absolute error on contigs with varying contamination levels (10\%-50\%). (C) Percentage of misidentified provirus sequences on contigs of purely viral origin. Y-axis presents the false positive rate, which is around 1\% for ViralQC.}
    \label{fig:contamination_predict_level}  
\end{figure}

\begin{figure*}[h!]
    \centering
    \includegraphics[width=0.98\linewidth]{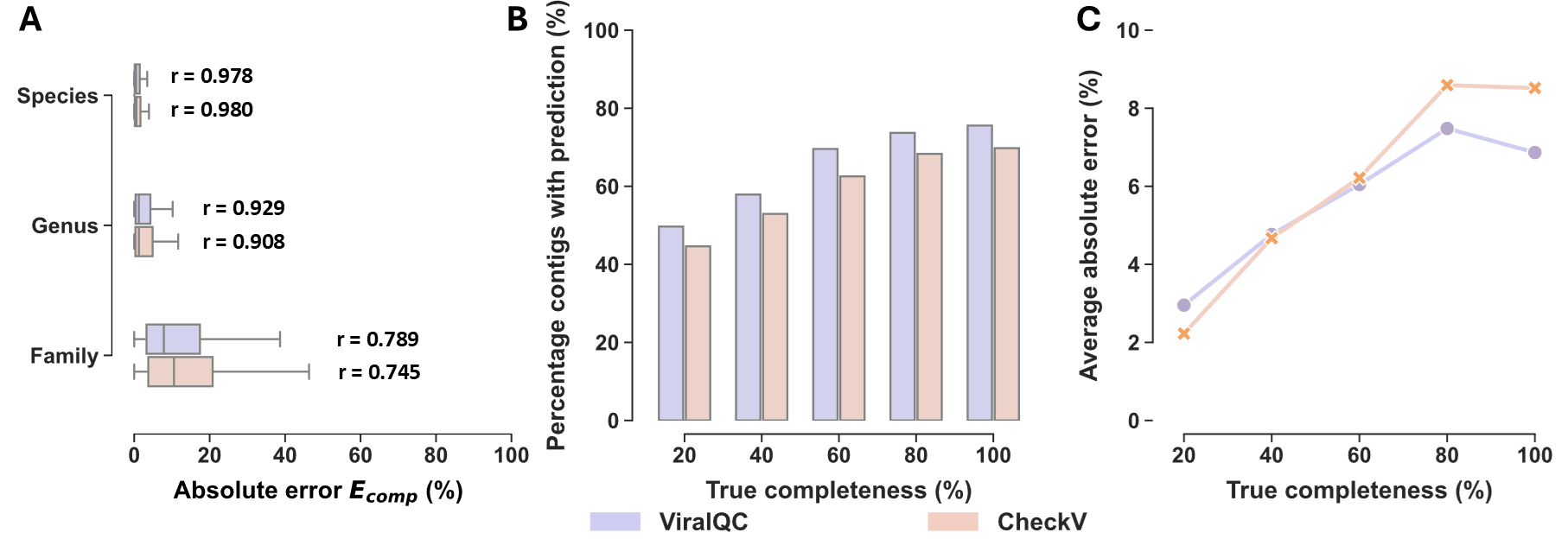}
    \caption{(A) The Performance of completeness estimation on leave-one-taxon-out datasets. Pearson's correlation coefficient is indicated by $r$. (B) The percentage of contigs estimated by each tool on contigs with completeness from 20\% to 100\%. \tool\ is able to estimate completeness for more inputs. (C) The average absolute error of estimated completeness on contigs with completeness from 20\% to 100\%. \tool\ provide more accurate results on contigs with completeness > 50\%.}
    \label{fig:completeness_percentage}  
\end{figure*}
\vspace{-0.5cm}
\subsubsection{Contamination detection result}

We evaluate contamination detection performance on the mock provirus dataset, which includes contigs of varying contig lengths (5k-20kbp) and contamination levels (10\%-50\%). First, we report the percentage of discovered proviruses in Fig. \ref{fig:contamination_predict_num}. Overall, \tool\ successfully identifies 86\% of proviruses, compared to 62\% detected by CheckV, highlighting \tool's superior sensitivity in detecting host contamination within provirus sequences. Next, the distribution of contaminations estimated by each tool against the true contamination is present in Fig. \ref{fig:contamination_predict_level} A. The results show that the lengths of contamination are correctly predicted by \tool\ across all contamination levels. Furthermore, Fig. \ref{fig:contamination_predict_level} B demonstrates that \tool\ precisely identifies the virus-host boundaries, with a median absolute error of only 3\%. In contrast, CheckV's predictions fluctuate significantly when host regions account for 50\% of the entire contig, indicating CheckV's difficulty in detecting accurate virus-host boundaries in sequences with large host regions due to the absence of marker genes. 

\paragraph{False positive rate} We also evaluate the performance of both tools on contigs of purely viral origin. Fig. \ref{fig:contamination_predict_level} C shows that \tool\ has a minimal tendency to misclassify purely viral sequences as contaminated, with a false positive rate as low as 1\%. This high specificity further solidifies \ the tool as a robust and reliable solution for identifying contamination.

\subsection{Completeness estimation}
Because completeness is calculated using a structure-aware similarity search against a reference set of complete viral genomes, we evaluate the performance of \tool\ in completeness estimation on test data with different similarities against the reference genomes.

\subsubsection{Leave-one-taxon-out dataset}

To construct test data with increased difficulty, we constructed benchmark datasets using a leave-one-taxon-out strategy. Complete viral assemblies were downloaded from the NCBI RefSeq database. After careful examination, incomplete sequences labeled as ``partial'' were removed. The resulting dataset comprised 13,715 assemblies and 17,007 viral genomes from 2,423 genera, representing a wide diversity of viral species. For genera containing more than one species, we randomly distribute the species to training and test sets in a ratio of nine to one. 
This leave-species-out strategy ensures a balanced distribution of viral taxa while ensuring that there is no overlap between training and test sets. Then, we applied the same strategy at the genus and family level, resulting in three separate datasets with increasing difficulty. To simulate realistic scenarios, complete genomes in these datasets were randomly fragmented to create contigs with completeness levels ranging from 20\% to 100\%.

\paragraph{Metrics}
We calculate the absolute error between the estimated completeness $\hat{y}$ and the true completeness $y$:
\begin{equation}
    \label{m2}
    {E}_{comp} = \left | y-\hat{y}  \right |
\end{equation}
For completeness estimation, we replace CheckV's complete genome database with the same database used by \tool\ to ensure a fair comparison.

\begin{figure}[b]
    \centering
    \vspace{-0.5cm}
    \includegraphics[width=0.9\linewidth]{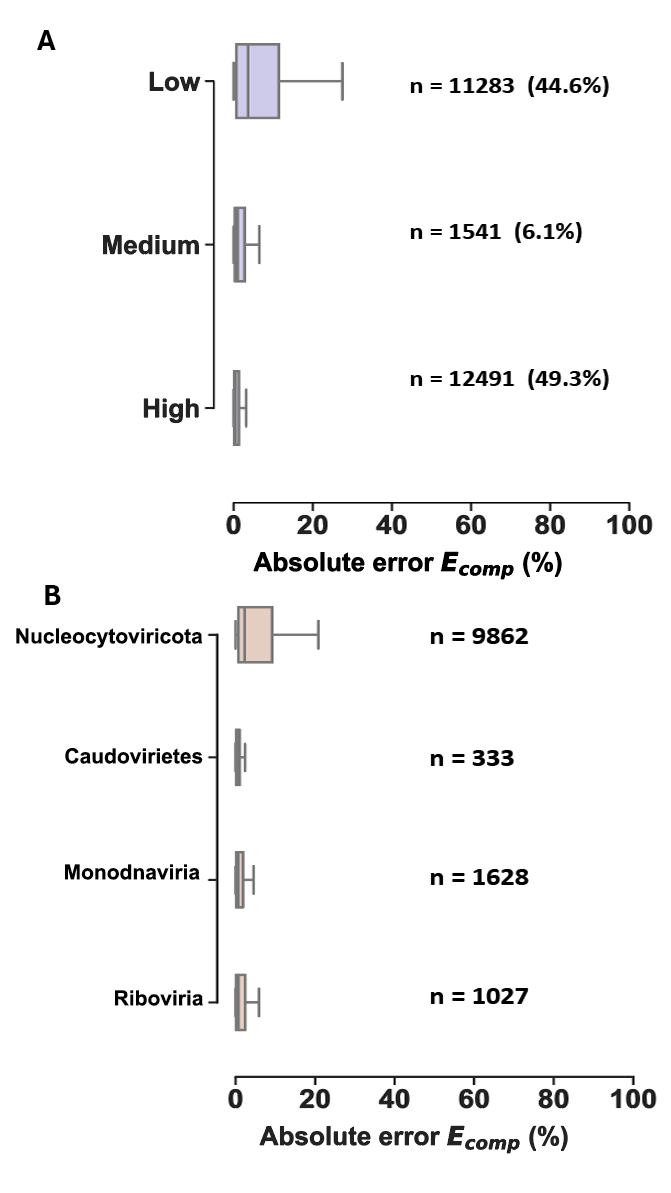}
    \vspace{-0.5cm}
    \caption{(A) \tool's performance on low-, medium- and high-confidence levels. (B) Medium- and high-confidence results of \tool\ on contigs from different viral groups. }
    \label{fig:complteness_confidence}  
\end{figure}

\subsubsection{Completeness estimation result}
We present the performance of completeness estimation using leave-one-taxon-out datasets, which simulate increasingly challenging scenarios by excluding certain taxonomic groups from the training set. As shown in Fig. \ref{fig:completeness_percentage} A, the mean unsigned error increases from 1.9\% at the species level to 13.1\% at the family level. This pattern reflects the greater genomic diversity observed at higher taxonomic levels, especially at the family level. Despite the challenge carried by the leave-family-out dataset, both \tool\ and CheckV achieve high performance at the species and genus levels, with Pearson's correlation coefficient (r) exceeding 0.9, indicating a strong relationship between the estimated and true completeness values.

We further compare their performance on contigs with varying completeness levels, ranging from 20\% to 100\%. As shown in Fig. \ref{fig:completeness_percentage} B, \tool\ is able to estimate completeness for a large proportion of contigs over all ranges, while CheckV fails to provide completeness results for certain inputs. Fig. \ref{fig:completeness_percentage} C further presents that although CheckV performs slightly better on short fragments with completeness < 50\%, \tool\ delivers more accurate results for medium- to high-quality (>50\% completeness) contigs. The results further highlight \tool\'s ability to effectively utilize the informative feature of protein organization for analyzing complete or near-complete contigs. This is particularly important in metagenomic studies, where high-quality contigs are often prioritized for downstream analyses. The ability of \tool\ to handle such contigs with greater precision makes it a more reliable tool for analyzing viral sequences.

\paragraph{Effective confidence prediction} While previous results include all the sequences, the estimated completeness on highly novel contigs may not be reliable. As a result, we report the results based on confidence levels. Fig. \ref{fig:complteness_confidence} A shows that the performance of \tool\ significantly improves on medium- and high-confidence predictions. Specifically, The mean estimation error decreases from 9\% to 1.5\%, and the standard deviation of the error is reduced from 14.3\% to 3.7\%, suggesting that the confidence tiers effectively distinguish reliable predictions. Such a mechanism is critical in real-world applications, where users can prioritize high-confidence results for downstream analyses and cautiously interpret lower-confidence predictions.

We also evaluate \tool's performance on four virus groups of main interest: dsDNA phages (\textit{Caudoviricetes}), NCLDVs (\textit{Nucleocytoviricota}), RNA viruses (\textit{Riboviria}), and ssDNA viruses (\textit{Monodnaviria}). The results with medium- and high-confidence in Fig. \ref{fig:complteness_confidence} B demonstrate that \tool\ achieves accurate completeness estimation across most groups, with an average error of 1.76\%.  However, the performance for Nucleocytoviricota stands out, with an average error of 7.2\%. The higher error can be attributed to the large variance in genome sizes within this group. For example, the genome size ranges from 120kb to 200kb within the family \textit{Ascoviridae}. Additionally, the limited availability of complete \textit{Nucleocytoviricota} genomes (162 in the NCBI RefSeq database) compared to Caudoviricetes (>5,000 genomes) highlights the need for more high-quality reference genomes to improve performance for this group.

\subsection{Results on real metagenomic data}

To evaluate \tool\ on real metagenomic sequencing data, we retrieved one dataset sequenced from \textit{Phragmites} plants rhizosphere soil (soil attached to plant roots) and bulk soil (nearby soil without plants) \cite{wang2021manganese}.

\begin{figure}[h!]
    \centering
    \includegraphics[width=0.9\linewidth]{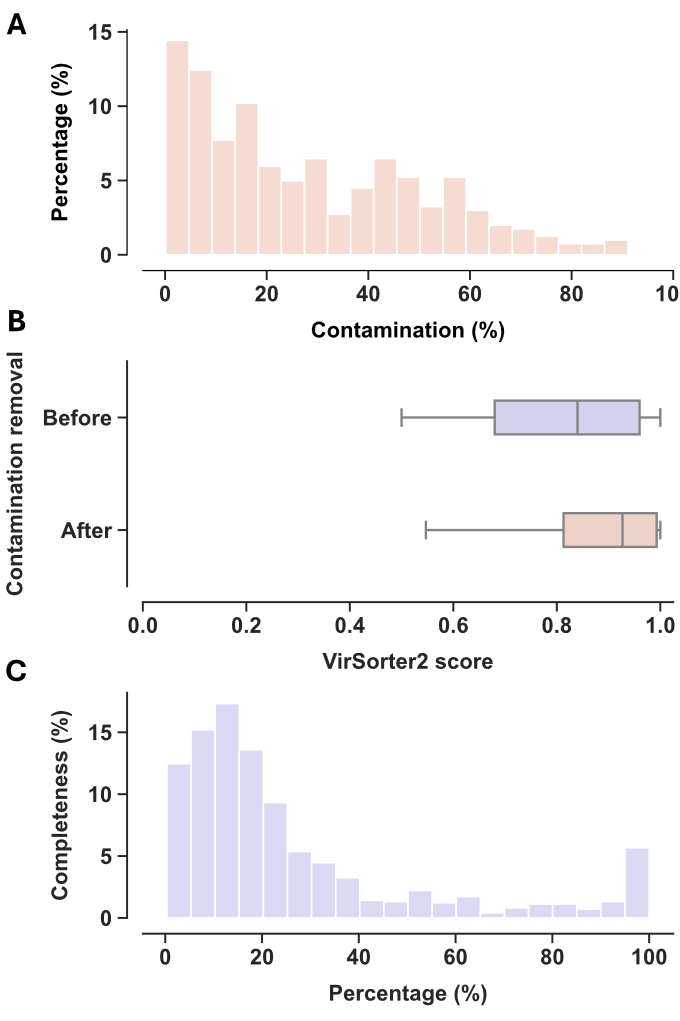}
    \caption{(A) The distribution of contamination estimated by \tool\ on VirSorter2-derived viral contigs from soil metagenome. (B) The comparison of viral score generated by VirSorter2 on predicted viral sequences before and after the removal of non-viral regions. (C) The distribution of completeness predicted by \tool\ with medium- and high-confidence on soil metagenome. }
    \label{fig:soil}  
    \vspace{-0.5cm}
\end{figure}

The raw sequencing reads are quality-controlled using fastp \cite{chen2018fastp} to remove low-quality reads and trim adapters. Then the resulting clean reads are assembled into contigs using Megahit \cite{li2015megahit}. Only contigs with lengths over 5kbp are reserved for analysis, resulting in a total of 461,310 contigs.
Then, we apply VirSorter2 to the assembled contigs and identify 5,331 viral contigs. Each contig is assigned a score indicating the likelihood of it being of viral origin. Then we employ \tool\ to assess the contamination within these viral contigs, specifically identifying non-viral regions. Contamination is detected in 402 contigs, with non-viral regions comprising more than 50\% of the entire sequence in 19\% of these cases (Fig. \ref{fig:soil} A). 
Such high contamination in viral contigs substantially impacts contig quality and may mislead downstream analyses such as taxonomic classification and functional annotation, emphasizing the critical role of contamination detection and removal in metagenomic studies, particularly in complex environments where microbial and viral sequences are highly intertwined. After removing the contamination, we re-evaluated the decontaminated contigs using VirSorter2. As shown in  Fig. \ref{fig:soil} B, the VirSorter2 scores increase significantly by 9\% after the removal of non-viral regions, proving that \tool\ can effectively improve the quality of viral contigs by identifying and mitigating contamination.

Furthermore, we assess the completeness of the viral contigs. Completeness is estimated with medium or high confidence for 18.5\% of the contigs. The distribution of completeness is displayed in Fig. \ref{fig:soil} C. Among these, 84\% are genome fragments (completeness < 50\%), while only 7\% are classified as complete or near-complete sequences (completeness $\ge$90\%). The results suggest that the majority of viral sequences in the rhizosphere soil metagenome are fragmented, likely due to the inherent complexity of the soil microbiome and the challenges associated with sequencing and assembling. Furthermore, the low proportion of complete or near-complete viral genomes highlights the need for additional refinement or complementary tools to improve genome recovery.

\section{Discussion}

In this work, we propose \tool, a tool designed for the quality assessment of metagenome-assembled viral contigs. \tool\ has two primary functions: detecting contamination within viral contigs and estimating their completeness. The major advancement in the contamination detection module stems from the adoption of foundation models. By leveraging both DNA and protein features, \tool\ can accurately identify virus-non-virus boundaries, enabling precise detection of non-viral regions. For completeness estimation, we implemented a novel graph-based approach that identifies the most related reference genome by analyzing synteny blocks and protein-based structure variants. Benchmarking on mock provirus dataset, leave-one-taxon-out dataset, and soil metagenomic sequencing data demonstrate that \tool\ achieves high sensitivity and specificity in contamination detection, along with accurate completeness estimation.

While \tool\ significantly enhances the quality assessment of viral sequences, there are several aims to optimize \tool\ in our future work. One possible improvement is to incorporate protein relationships within the same contig to refine the contamination detection performance. As stated in \cite{guan2024phago}, nearby proteins may share related functions or work together to perform specific tasks, offering valuable insights into the genome context. Leveraging these relationships could enhance the \tool's ability to detect contamination more comprehensively. Another promising direction is to expand the quality assessment framework by incorporating additional metrics such as the presence of core mark genes or auxiliary metabolic genes, providing a more holistic evaluation of viral contigs. These enhancements would further solidify \tool's role as a robust tool for analyzing metagenomic viral sequences.

%
%


\section*{Funding}
This work was supported by the Hong Kong Research Grants Council (RGC) General Research Fund (GRF) [11209823], the Hong Kong Innovation and Technology Fund (ITF) [MRP/071/20X], and the City University of Hong Kong.

\bibliographystyle{unsrt}  
\bibliography{references}

\end{document}